\documentclass[aps,prl,preprint,showpacs,groupedaddress]{revtex4-2}
\usepackage{graphicx,units,amsmath,amssymb,bm}
\usepackage{yhmath}

\graphicspath{{fig/}}

\newlength{\gw}

\begin{document}

\setlength{\gw}{0.72\linewidth}

\title{Symmetry-Protected Topological relationship between $SU(3)$ and
  $SU(2)\times{U(1)}$ in Two Dimension}

\author{Ning Wang}

\author{Qiao Zhuang}
\email[]{zhuangqiao@sdjzu.edu.cn}%


\affiliation{School of Science, Shandong Jianzhu University,
  1000 Fengming Road, Jinan, Shandong 250101, China}

\date{\today}

\begin{abstract}
  Symmetry-protected topological $\left(SPT\right)$ phases are gapped
  short-range entangled states with symmetry $G$, which can be
  systematically described by group cohomology theory. $SU(3)$ and
  $SU(2)\times{U(1)}$ are considered as the basic groups of Quantum
  Chromodynamics and Weak-Electromagnetic unification, respectively.
  In two dimension $(2D)$, nonlinear-sigma models with a quantized
  topological Theta term can be used to describe nontrivial SPT
  phases. By coupling the system to a probe field and integrating out
  the group variables, the Theta term becomes the effective action of
  Chern-Simons theory which can derive the response current density.
  As a result, the current shows a spin Hall effect, and the quantized
  number of the spin Hall conductance of SPT phases $SU(3)$ and
  $SU(2)\times{U(1)}$ are same. In addition, relationships between
  $SU(3)$ and $SU(2)\times{U(1)}$ which maps $SU(3)$ to $SU(2)$ with a
  rotation $U(1)$ will be given.
\end{abstract}


\maketitle

In condensed matter physics, Quantum Chromodynamics $\left(QCD\right)$
is a proper gauge theory to describe Strong interaction by
investigating the relationship between the basic element
$\left(quark\right)$ and the gauge
field~\cite{Shifman-ma1979A,Shifman-ma1979B}. Gapped phases of quantum
matter are naturally described by topological quantum field theories
$\left(TQFTs\right)$ at low energy and long
distance~\cite{Wang-qr2019}. In two dimension $\left(2D\right)$,
Abelian and non-Abelian Chern-Simons theories can be used to capture
the topological properties of fractional quantum Hall
conductance~\cite{Wen-xg2004,Nayak-c2008}. And in the TQFTs, there is
another interesting method, symmetry-protected topological
$\left(SPT\right)$ phases~\cite{Ye-p2016,Gu-zc2014,Chen-x2013}, which
can be transformed to product states via local unitary
$\left(LU\right)$
transformations~\cite{Vidal-g2007,Levin-ma2005,Verstraete-f2005}. In
reference~\cite{Chen-x2011,Chen-x2013}, the results for
one-dimensional SPT phase are generalized to any dimensions. So far,
the model of quark~\cite{Callan-cg1976,Poggio-ec1976,Witten-e1979},
construction of baryon~\cite{Chodos-a1974,Loffe-bl1981}, and
normalized field~\cite{Singer-m1980,Weinberg-s1972,Corrigan-e1976} are
constructed out of group $SU(3)$. The Weinberg-Salam model is a proper
model with combinatorial group $SU(2)\times{U(1)}$, whose theoretical
calculation results are consistent with the experiment(i.e. boson
$W^{+}$, $W^{-}$ and $Z^{0}$ get quality, photons are
massless)~\cite{Buras-aj1978,Dimopoulos-s1981,Luo-lf1980}. This model
can be used to describe Weak-Electromagnetic interaction, which
consists of Weak interaction normalized group $SU(2)$ and the
Electromagnetic interaction normalized group $U(1)$. If we split group
$SU(3)$ into the product of two more basic groups as
$SU(2)\times{U(1)}$, the QCD will become a new version. Using this
method, some similar properties between group $SU(3)$ and
$SU(2)\times{U(1)}$ can be easily founded.

In the work present here, we will introduce two kind of SPT
phases---$SU(3)$ and $SU(2)\times{U}(1)$. If these phases couple to
external probe field individually, the derived results will show that
the spin Hall conductance are quantized, and the quantized number of
$SU(3)$ and $SU(2)\times{U}(1)$ are same.

Principal chiral nonlinear sigma model $\left(PCM\right)$ with a Theta
term, which has action as~\cite{Xu-c2013}

\begin{equation}
  \label{eq:eq01}
  S=\int_{M}d{\tau}d^{2}x\frac{1}{Q}Tr\left[\left(g^{-1}{\partial}_{\mu}g\right)\left(g^{-1}{\partial}_{\mu}g\right)\right]
  +\frac{i\Theta}{24{\pi}^2}Tr\left[{\varepsilon}^{\mu\nu\lambda}\left(g^{-1}{\partial}_{\mu}g\right)\left(g^{-1}{\partial}_{\mu}g\right)\left(g^{-1}{\partial}_{\mu}g\right)\right]
\end{equation}

\noindent where the second term on the right-hand side is the action
of topological term, $g$ is a group element of $SU(n)$, $M$ the
Euclidian space-time manifold, and ${\Theta}=2{\pi}{k}$ with
$k\in{\mathbb{Z}}$ meaning PCM is quantized. When ${\Theta}=\pi{k}$,
the system also has two discrete symmetries, reflection varies as
$x\rightarrow{-x}$ and time reversal varies as $i\rightarrow{-i}$,
$t\rightarrow{-t}$(consequently $\tau\rightarrow{\tau}$). It is easy
to check that reflection will not affect the quantized number.
According to~\cite{Liu-zx2013}, the discussion about the time reversal
shows that the quantized number will not change, neither. As we can
see from the equation(\ref{eq:eq01}), if $Q$ flows to infinity, the
action will flow to a fixed point where only the topological term
remains. In spite of ignoring the first term of
equation(\ref{eq:eq01}), the physical properties of the system will
not change. In the rest part of this letter, gauge symmetry will be
discussed in the fixed point condition.

Following the research~\cite{Liu-zx2013}, equation(\ref{eq:eq01}) is
verified to be invariant under a symmetry
$SU(n)_{L}\times{SU}(n)_{R}$, where $SU(n)_{L}$ and $SU(n)_{R}$ are
left and right symmetry group, respectively. The group element $g$
varies as $g\rightarrow{hg}$ for $h\in{SU}(n)_{L}$, while varies as
$g\rightarrow{gh^{-1}}$ for $h\in{SU}(n)_{R}$. It is easy to check
that the equation(\ref{eq:eq01}) is invariable under symmetry group
$SU(n)_{L}$, but no more invariant under symmetry group $SU(n)_{R}$.

In order to investigate the gauge symmetry of PCM, the Theta term
should be couple to an external probe field ${A}$ by replacing every
$g^{-1}{\partial}_{\mu}g$ term with
$g^{-1}\left({\partial}_{\mu}+{A}_{\mu}\right)g$.

At the fixed point, Theta term becomes

\begin{equation}
  \label{eq:eq02}
  \frac{\Theta}{24{\pi}^2}\int_{M}Tr\left[g^{-1}\left(d+A\right)g\right]^3
  =\frac{\Theta}{24{\pi}^2}\int_{M}Tr\left[\left(g^{-1}dg\right)^3+A^3\right.
  \left.+3dgg^{-1}{\wedge}F+3d\left(dgg^{-1}{\wedge}A\right)\right]
\end{equation}

\noindent where $F=dA+A\wedge{A}$ is the field strength of the probe
field $A$.  Focusing on the four terms on the right-hand side of the
equation(\ref{eq:eq02}), we can obtain that the first term is the
Theta term in equation(\ref{eq:eq01}), the second term is the pure
probe field function, and the rest two terms are functions of $A$ and
$dgg^{-1}$. For simplicity, the Theta term can be reduced to the
effective field theory of the external field $A$ by integrating out
the group variables $g$. the effective action of probe field $A$ can
be expressed as the Chern-Simons action

\begin{equation}
  \label{eq:eq03and04and09}
  S_{\textnormal{eff}}\left(A\right)=i\frac{\Theta}{8{\pi}^2}\int_{M}Tr\left(A{\wedge}F-\frac{1}{3}A^3\right)
  =i\frac{\Theta}{16{\pi}^2}\int_{M}{\varepsilon}^{\mu\nu\lambda}Tr\left(A^{a}_{\mu}\partial_{\nu}A^{a}_{\lambda}+\frac{2}{3}{\varepsilon}_{abc}A^{a}_{\mu}A^{b}_{\nu}A^{c}_{\lambda}\right)
\end{equation}

\noindent where the construction of $A$ depends on the different
conditions, which will be discussed in the following sections.

Firstly, let us talk about the $SU(3)$ group. According to the
reference~\cite{Chen-x2013}, the $SU(3)$ SPT phases can be classified
by group cohomology class
$\mathcal{H}^{3}\left[SU(3),U(1)\right]=\mathbb{Z}$ in $2D$(the
details of the calculation can be founded in the Supplement Material
Section A). This non-abelian probe field can be expressed as
$A_{\mu}=\sum_{a}{A^{a}_{\mu}}{T^{a}}=\frac{1}{2}\sum_{a}{A^{a}_{\mu}}\left(2T^{a}\right)$,
where $2T^{a}$ are the eight Gell-Mann matrixes that generate the
Quantum Chomodynamics theory. The trace
$Tr\left(T^{a}{T}^{b}\right)={\frac{1}{2}{\delta}^{ab}}$ contributes
an extra coefficient $\frac{1}{2}$ in the
equation(\ref{eq:eq03and04and09}).
 
By calculating the variation of equation(\ref{eq:eq03and04and09}), the
response current expressed as following

\begin{equation}
  \label{eq:eq05}
  \mathcal{J}^{a}_{\mu}=\frac{{\delta}S_{\textnormal{eff}}}{{\delta}A^{a}_{\mu}}=i\frac{\Theta}{8{\pi}^2}{\varepsilon}^{\mu\nu\lambda}\left({\partial}_{\nu}A^{a}_{\lambda}+{\varepsilon}_{abc}A^{b}_{\nu}A^{c}_{\lambda}\right)
\end{equation}

Without loss of generality, we assume that the probe field $A$ only
contains $A^{a}$ component. Then the time component of the current in
equation(\ref{eq:eq05}) can be expressed as

\begin{equation}
  \label{eq:eq06}
  \mathcal{J}^{a}_{t}=i\frac{\Theta}{8{\pi}^2}\left({\partial}_{x}A^{a}_{y}-{\partial}_{y}A^{a}_{x}\right)
\end{equation}

\noindent and the space components can be expressed as

\begin{equation}
  \label{eq:eq071}
  \mathcal{J}^{a}_{x}=i\frac{\Theta}{8{\pi}^2}\left({\partial}_{y}A^{a}_{t}-{\partial}_{t}A^{a}_{y}\right)
\end{equation}

\begin{equation}
  \label{eq:eq072}
  \mathcal{J}^{a}_{y}=i\frac{\Theta}{8{\pi}^2}\left({\partial}_{t}A^{a}_{x}-{\partial}_{x}A^{a}_{t}\right)
\end{equation}

Equation(\ref{eq:eq06})-(\ref{eq:eq072}) can be used to express the
spin Hall effect, if the time and space components of the current
representations are considered as magnetic and electric field,
respectively. The spin Hall conductance is quantized as
$\frac{\Theta}{8{\pi}^2}$.

Secondly, in $2D$, $SU(2)\times{U}(1)$ SPT phases are classified by
group cohomology class
$\mathcal{H}^{3}\left[SU(2)\times{U}(1),U(1)\right]=\mathbb{Z}$(For
details, see the Supplement Material Section B). At the fixed point,
the Theta term of the PCM of $SU(2)\times{U}(1)$ becomes

\begin{equation}
  \label{eq:eq08}
  \begin{aligned}
  S&=\frac{i\Theta}{24{\pi}^2}Tr\left\{{\varepsilon}^{\mu\nu\lambda}\left[{\left(gh\right)}^{-1}{\partial}_{\mu}{\left(gh\right)}\right]\left[{\left(gh\right)}^{-1}{\partial}_{\nu}{\left(gh\right)}\right]\left[{\left(gh\right)}^{-1}{\partial}_{\lambda}{\left(gh\right)}\right]\right\}\\
  &=\frac{i\Theta}{24{\pi}^2}Tr\left\{{\varepsilon}^{\mu\nu\lambda}\left[\left({g}^{-1}\partial_{\mu}{g}\right)\left({g}^{-1}\partial_{\nu}{g}\right)\left({g}^{-1}\partial_{\lambda}{g}\right)+\left({h}^{-1}\partial_{\mu}{h}\right)\left({h}^{-1}\partial_{\nu}{h}\right)\left({h}^{-1}\partial_{\lambda}{h}\right)\right]\right\}
  \end{aligned}
\end{equation}

\noindent where $g\in{SU(2)}$ and $h\in{U(1)}$. The two terms on
right-side of second row represent the action of group $SU(2)$ and
group $U(1)$, respectively. For group $SU(2)$, its construction can be
referred to the same part of $SU(3)$ as
equation(\ref{eq:eq03and04and09}), however probe fields satisfy
$A_{\mu}=\sum_{a}{A^{a}_{\mu}}{T^{a}}=\frac{1}{2}\sum_{a}{A^{a}_{\mu}}\left(2T^{a}\right)$,
where $2T^{a}$ are three Pauli matrixes. Do the same derivation as
$SU(3)$, the time and space components of current can also be
expressed as equations(\ref{eq:eq06})-(\ref{eq:eq072}). So the
quantized number of spin Hall effect is $\frac{\Theta}{8{\pi}^2}$. It
is noted that $U(1)$ is an Abelian group and its construction is
different from $SU(n)$. According to the
reference~\cite{Dunne-gv1999}, Abelian version of the Chern-Simons
Lagrangian is

\begin{equation}
  \label{eq:eq13}
  \mathcal{L}_{\textnormal{CS}}=i\frac{\Theta}{8{\pi}^2}{\varepsilon}^{\mu\nu\lambda}Tr{A_{\mu}}{\partial}_{\nu}A_{\lambda}
\end{equation}

Then the effective action and the response current density become

\begin{equation}
  \label{eq:eq14}
  S_{\textnormal{eff}}\left(A\right)=i\frac{\Theta}{8{\pi}^2}\int_{M}{\varepsilon}^{\mu\nu\lambda}TrA_{\mu}{\partial}_{\nu}A_{\lambda}
\end{equation}

\begin{equation}
  \label{eq:eq15}
  \mathcal{J}^{\nu}_{\mu}=\frac{{\delta}S_{\textnormal{eff}}}{{\delta}A_{\mu}}=i\frac{\Theta}{8{\pi}^2}{\varepsilon}^{\mu\nu\lambda}Tr{\partial}_{\nu}A_{\lambda}
\end{equation}

Combine equation(\ref{eq:eq05}) and (\ref{eq:eq15}), the response
current density of the group $SU(2)\times{U}(1)$ is

\begin{equation}
  \label{eq:eq16}
  \mathcal{J}^{a}_{\mu}=i\frac{\Theta}{8{\pi}^2}{\varepsilon}^{\mu\nu\lambda}Tr\left({\partial}_{\nu}A^{a}_{\lambda}+{\varepsilon}_{abc}{A}^{b}_{\nu}{A}^{c}_{\lambda}+{\partial}_{\nu}{B}_{\lambda}\right)
=i\frac{\Theta}{8{\pi}^2}{\varepsilon}^{\mu\nu\lambda}Tr\left[{\partial}_{\nu}\left({A}^{a}_{\lambda}+{B}_{\lambda}\right)+{\varepsilon}_{abc}{A}^{b}_{\nu}{A}^{c}_{\lambda}\right]
\end{equation}

\noindent where $A^{a}_{\lambda}\in{SU(2)}$ and
$B_{\lambda}\in{U(1)}$. Assuming ${A}^{a}_{\lambda}+{B}_{\lambda}$ is
a new probe field, it is easy to find that the response density
construction of group $SU(2)\times{U(1)}$ is similar to group $SU(3)$.
The group $SU(3)$ is non-Abelian and satisfies the commutation
relations
$\left[{T}^{a},{T}^{b}\right]={\mathnormal{f}_{abc}{T}^{c}}$, where
$\mathnormal{f}_{abc}$ is the structural constant. The basic group of
$SU(2)\times{U(1)}$ satisfies
$\left[{T}^{a},{T}^{b}\right]={\mathnormal{g}_{abc}{T}^{c}}$. The structural constant $\mathnormal{g}_{abc}$ and $\mathnormal{f}_{abc}$
are similar but not the same. Therefore, there should be a potential
relationship between group $SU(3)$ and $SU(2)\times{U(1)}$.

In order to obtain the relationship, we give Gell-Mann matrixes and
Pauli matrixes as following

\begin{equation}
\centering
\label{eq:eq17}
\begin{aligned}
{\lambda}_{1}=\left(
  \begin{array}{ccc}
    0 & 1 & 0 \\
    1 & 0 & 0 \\
    0 & 0 & 0 \\
  \end{array}
\right),
{\lambda}_{2}=\left(
  \begin{array}{ccc}
    0 & -i & 0 \\
    i & 0 & 0 \\
    0 & 0 & 0 \\
  \end{array}
\right),
{\lambda}_{3}=\left(
  \begin{array}{ccc}
    1 & 0 & 0 \\
    0 & -1 & 0 \\
    0 & 0 & 0 \\
  \end{array}
\right),
{\lambda}_{4}=\left(
  \begin{array}{ccc}
    0 & 0 & 1 \\
    0 & 0 & 0 \\
    1 & 0 & 0 \\
  \end{array}
\right),\\\\
{\lambda}_{5}=\left(
  \begin{array}{ccc}
    0 & 0 & -i \\
    0 & 0 & 0 \\
    i & 0 & 0 \\
  \end{array}
\right),
{\lambda}_{6}=\left(
  \begin{array}{ccc}
    0 & 0 & 0 \\
    0 & 0 & 1 \\
    0 & 1 & 0 \\
  \end{array}
\right),
{\lambda}_{7}=\left(
  \begin{array}{ccc}
    0 & 0 & 0 \\
    0 & 0 & -i \\
    0 & i & 0 \\
  \end{array}
\right),
{\lambda}_{8}=\frac{1}{\sqrt{3}}\left(
  \begin{array}{ccc}
    1 & 0 & 0 \\
    0 & 1 & 0 \\
    0 & 0 & -2 \\
  \end{array}
\right)
\end{aligned}
\end{equation}

\noindent and

\begin{equation}
\label{eq:eq19}
{\sigma}_{1}=\left(
  \begin{array}{cc}
    0 & 1 \\
    1 & 0  \\
  \end{array}
\right),
{\sigma}_{2}=\left(
  \begin{array}{cc}
    0 & -i \\
    i & 0 \\
  \end{array}
\right),
{\sigma}_{3}=\left(
  \begin{array}{cc}
    1 & 0  \\
    0 & -1 \\
  \end{array}
\right).
\end{equation}

Then, assuming group $SU(2)$ maps to $SU(3)$

\begin{equation}
\begin{aligned}
\label{eq:eq20}
&{\sigma}_{1}\rightarrow{\lambda}_{1},{\lambda}_{4},{\lambda}_{6},\\
&{\sigma}_{2}\rightarrow{\lambda}_{2},{\lambda}_{5},{\lambda}_{7},\\
&{\sigma}_{3}\rightarrow{\lambda}_{3},{\lambda}_{8}.
\end{aligned}
\end{equation}

\begin{figure}
  \centering
  \includegraphics[width=\linewidth]{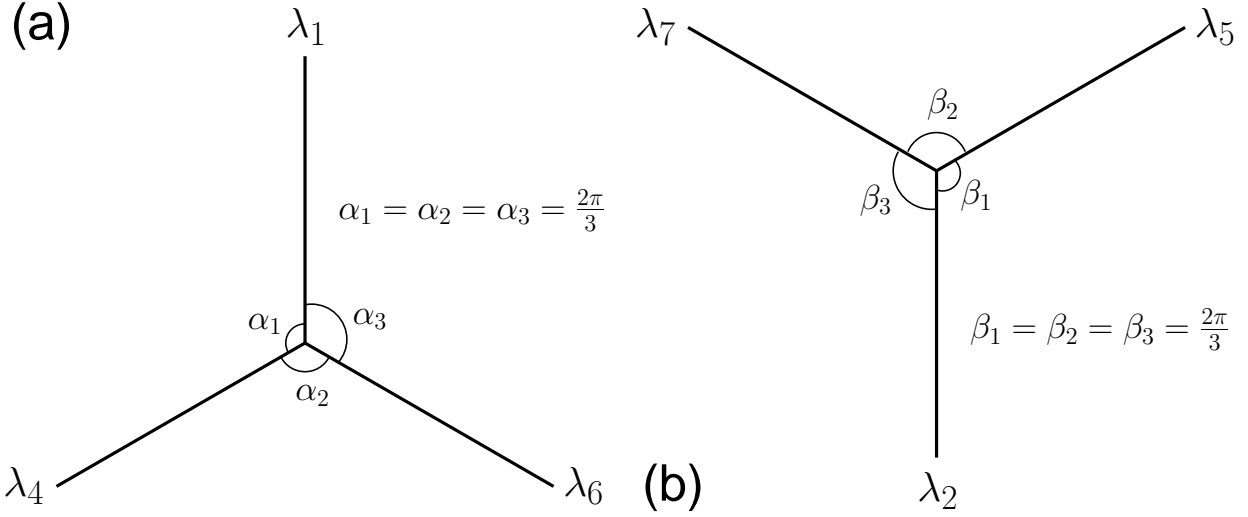}
  \caption{In $2D$, $U(1)$ rotation relationship between the matrixes
    of $SU(3)$. (a) matrixes $\lambda_{1}$, $\lambda_{4}$ and
    $\lambda_{6}$ form a new group, every vertex rotates to another
    by $\frac{2\pi}{3}$. (b) matrixes $\lambda_{2}$, $\lambda_{5}$ and
    $\lambda_{7}$ form a subset, every vertex rotates to another by
    $\frac{2\pi}{3}$.}
  \label{fig:lambda_fig}
\end{figure}

\begin{figure}
  \centering
  \includegraphics[width=\linewidth]{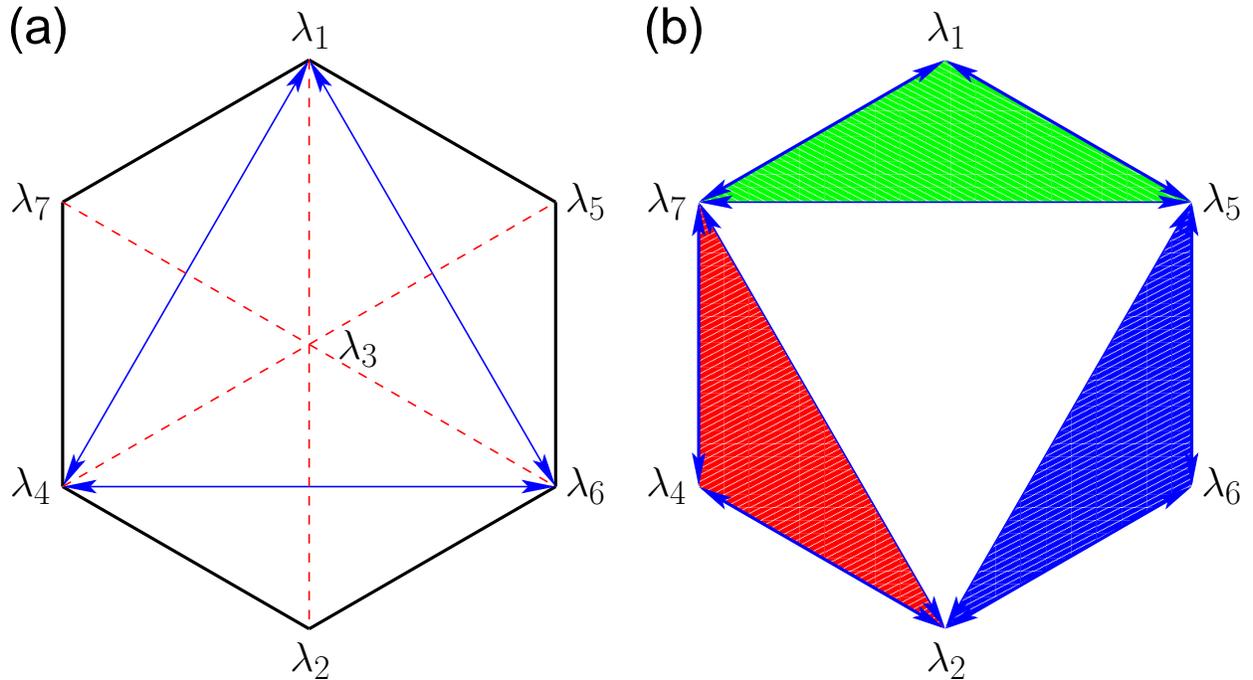}
  \caption{(color online). Every two vertexes have a relationship. The start vertex of the solid arrow is the matrix which can construct a new one; the end vertex is the matrix which is constructed by another. (a) Dotted line is the anticommute relationships between $\lambda_3$ and other, and two endpoints of dotted line across $\lambda_3$ anticommute each other. Every vertex in triangle $\lambda_1$,$\lambda_4$,$\lambda_6$ has two incoming edges and two outgoing edges, which makes $\lambda_1$,$\lambda_4$,$\lambda_6$ be a new group. (b) The red shaded area shows $\lambda_2$,$\lambda_4$,$\lambda_7$ construct a group; the green shaded area shows $\lambda_1$,$\lambda_5$,$\lambda_7$ construct a group; the blue shaded area shows $\lambda_2$,$\lambda_5$,$\lambda_6$ construct a group.}
  \label{fig:lambda_fig1}
\end{figure}

Picking some matrixes from the group $SU(3)$, it is clear that there
are some intrinsic relationships among them. The maps
in(\ref{eq:eq20}) give three relationships. $\lambda_1$, $\lambda_4$
and $\lambda_6$ in the first relationship can be treated as the
$3\times3$ external matrixes of $\sigma_1$. These matrixes satisfy the
anticommutation relationship
$\left\{\lambda_i,\lambda_j\right\}=\lambda_k$(where
$i$,$j$,$k$=$1$,$4$,$6$), which construct a new group.  Considering
matrixes $\lambda_1$, $\lambda_4$ and $\lambda_6$ are all off-diagonal
and their elements are symmetrical about the diagonal, we take group
$U(1)$ as a rotation that rotates matrixes diagonally.  Subsets of
group $SU(3)$ by the rotation $U(1)$ can be shown in
Fig~\ref{fig:lambda_fig}. The second relationship contains
$\lambda_2$, $\lambda_5$ and $\lambda_7$.  Unfortunately, there is
neither commutation nor anticommutation between $\lambda_2$,
$\lambda_5$ and $\lambda_7$, in spite of they seem as the $3\times3$
external matrixes of $\sigma_2$. If $\lambda_2$, $\lambda_5$ and
$\lambda_7$ are treated as the subset of $SU(3)$, rotation $U(1)$ will
still works in structure(Fig~\ref{fig:lambda_fig}(b)). The last
relationship contains $\lambda_3$ and $\lambda_8$, which are invariant
by the rotation $U(1)$ because they are both diagonal. For
$\lambda_3$, from the anticommutation relationship between $\lambda_3$
and other matrixes of $SU(3)$ except $\lambda_8$,

\begin{equation}
  \label{eq:eq21}
  \begin{aligned}
    \left\{\lambda_3,\lambda_1\right\}=0,
    \left\{\lambda_3,\lambda_2\right\}=0,
    \left\{\lambda_3,\lambda_4\right\}=\lambda_4,\\
    \left\{\lambda_3,\lambda_5\right\}=\lambda_5,
    \left\{\lambda_3,\lambda_6\right\}=-\lambda_6,
    \left\{\lambda_3,\lambda_7\right\}=-\lambda_7
  \end{aligned}
\end{equation}

\noindent there should be a relationship between every two
matrixes(Fig~\ref{fig:lambda_fig1}). However, matrix except
$\lambda_8$ anticommutes $\lambda_8$ will obtain $2$ times or negative
of the matrix itself,

\begin{equation}
  \label{eq:eq22}
  \begin{aligned}
    \left\{\lambda_8,\lambda_1\right\}=2\lambda_1,
    \left\{\lambda_8,\lambda_2\right\}=2\lambda_2,
    \left\{\lambda_8,\lambda_3\right\}=2\lambda_3,
    \left\{\lambda_8,\lambda_4\right\}=-\lambda_4,\\
    \left\{\lambda_8,\lambda_5\right\}=-\lambda_5,
    \left\{\lambda_8,\lambda_6\right\}=-\lambda_6,
    \left\{\lambda_8,\lambda_7\right\}=-\lambda_7
  \end{aligned}
\end{equation}

\noindent in this way, $\lambda_8$ can be considered as a constraint
condition which makes the eight matrixes form a group completely.

In summary, we study PCM actions of $SU(3)$ and $SU(2)\times{U(1)}$,
which both have a Theta term. Coupling the system to a probe field and
integrating out the group variables, the results can be considered as
effective action of Chern-Simons theory. As a consequence, the spin
Hall conductance of $SU(3)$ and $SU(2)\times{U(1)}$ both are quantized
as $\frac{\Theta}{8{\pi}^2}$. Furthermore, in $SU(3)$, we calculate
anticommutation relationships which show that there is an intrinsic
connection between every two matrixes. In order to simplify the
$SU(3)$, we give a mapping from $SU(2)$ to $SU(3)$ under the rotation
$U(1)$. This rotation rotates matrixes diagonally, which classifies
$SU(3)$ into three categories. Every category is invariant under the
rotation $U(1)$, which makes a simpler model $SU(2)\times{U(1)}$ to
represent $SU(3)$. In addition, there are three more subgroups in
$SU(3)$ by the anticommutation relationships, which can be used to
investigate the intrinsic constructions and properties of the $quark$.
Our work might be useful to investigate the further significant
information of QCD by above method. The current results do work in
$2D$ condition, the work in higher dimension will be carried out in
future.

This work is supported by the National Natural Science Foundation of
China(Project No. 11374193).

\providecommand{\noopsort}[1]{}\providecommand{\singleletter}[1]{#1}%

\end{document}